\documentclass[twocolumn,preprintnumbers,nofootinbib,aps,prd,floatfix]{revtex4}
\pdfoutput=1 % if your are submitting a pdflatex to JHEP (i.e. if you have images in pdf, png or jpg format)

\usepackage{subfigure,graphicx,amsmath,amssymb,hyperref}

\newcommand{\bea}{\begin{eqnarray}}
\newcommand{\eea}{\end{eqnarray}}

\def\beq#1\eeq{\begin{align}#1\end{align}}
\def\beqnn#1\eeq{\begin{align*}#1\end{align*}}

\def \tev {\text{ TeV}}
\def \gev {\text{ GeV}}
\def\ifb{\text{fb$^{-1}$}}

\usepackage[usenames,dvipsnames]{xcolor}
\definecolor{darkgreen}{rgb}{0,0.5,0}

\newcommand{\rkks}{$R_{K^{(*)}}$}

\newcommand{\rds}{$R_{D^*}$}
\newcommand{\rdds}{$R_{D^{(*)}}$}
\newcommand{\rjp}{$R_{J/\psi}$}

\begin{document}

\preprint{UCI-HEP-TR 2018-03}
\title{$B$ Anomalies and Leptoquarks at the LHC: Beyond the Lepton-Quark Final State}

\author{Angelo Monteux}
\email{monteuxa@uci.edu}
\author{Arvind Rajaraman}
\email{arajaram@uci.edu}
\affiliation{Department of Physics and Astronomy,\\
University of California, Irvine, CA 92697-4575 USA
}

\begin{abstract}
Leptoquarks provide some of the simplest explanations to the hints of lepton flavor non-universality in $B$ decays.
In particular, a new confining gauge group can provide a natural and appealing origin for the leptoquarks.
So far, direct collider searches have been based on two body decays, namely to a quark and a lepton. 
We study how composite dynamics can give rise to additional states resulting in modified collider signatures of leptoquarks, as well as new production modes in cascade decays of heavier states. Instead of the standard signature, each leptoquark can result in as many as four jets and a lepton.
We reinterpret relevant ATLAS and CMS searches to set limits on this scenario, show how this can relax the current bounds, and propose ways to better constrain this class of models in the future. For example, we show that a leptoquark related to the $R_{D^{(*)}}$ anomaly could still be as light as 500\gev.
\end{abstract}

\maketitle

%%%%%%%%%%%%%%%%%%%%%%%%%%%%%%%%%%%%%%%%%%%%%%%%%%%

\section{Introduction}

Some of the most long-standing hints of Physics beyond the Standard Model (SM) are found
 in decays of $B$ mesons \cite{Lees:2013uzd,Aaij:2014ora,Huschle:2015rga,Aaij:2015yra,Hirose:2016wfn,Aaij:2017deq,Aaij:2017vbb,Aaij:2017tyk}. In particular, multiple independent measurements at Babar, Belle 
and LHCb have been made that are in tension with lepton flavor universality,  the principle that SM weak bosons couple universally to leptons of different generations. These include discrepancies 
in the charged current process $B\to D^{(*)}\ell\nu$, which in the Standard Model is mediated 
by the $W$ at tree level, as well as the neutral current process $B\to K^{(*)}\ell^+\ell^-$ (which is 
loop-level in the SM).

\begin{table*}[tb]
\begin{center}
$$
\begin{array}{c|c|cc|cc|c}
\text{Name} & \text{Definition} & \text{Observed value}&\text{ [ref]} & \text{SM prediction} & \text{[ref]} & \text{Discrepancy}\\\hline
R_{K} & \frac{Br(B^+\rightarrow K^+\mu^+\mu^-)}{ Br(B^+\rightarrow K^+ e^+e^-)} & 0.745^{+0.090}_{-0.074}\pm 0.036 &\text{\cite{Aaij:2014ora}}& 1& \text{\cite{Hiller:2003js,flavio}} &2.6\sigma
\\
R_{K^{*}} & \frac{Br(B^0\rightarrow K^{0*}\mu^+\mu^-)}{ Br(B^0\rightarrow K^{0*} e^+e^-)} & [0.66, 0.69]^{+0.11}_{-0.07}\pm0.03 &\text{\cite{Aaij:2017vbb}} & [0.926, 0.9965]\pm0.0005& \text{\cite{Hiller:2003js,flavio}} & [2.2\sigma, 2.5\sigma]
\\\hline
R_{D} & \frac{Br(B\rightarrow D^{*}\tau^-\nu)}{ Br(B\rightarrow D \ell^-\nu)} & 0.407 \pm 0.039 \pm 0.024  &\text{\cite{hflav2017}}&  0.299 \pm 0.011 &\text{\cite{Lattice:2015rga}} & 2.3\sigma
\\
R_{D^{*}} & \frac{Br(B\rightarrow D^{*}\tau^-\nu)}{ Br(B\rightarrow D^{*} \ell^-\nu)} & 0.304 \pm 0.013 \pm 0.007 &\text{\cite{hflav2017}} & 0.252\pm0.003 & \text{\cite{Fajfer:2012vx}} & 3.4\sigma
\\
R_{J/\psi}  & \frac{Br(B_c^+\rightarrow J/\psi \tau^+\nu)}{ Br(B_c^+\rightarrow J/\psi \mu^+\nu)} & 0.71\pm 0.17\pm0.18 &\text{\cite{Aaij:2017tyk}}&0.29\pm 0.07 & \text{\cite{Wen-Fei:2013uea}} &1.7\sigma
\end{array}
$$
\end{center}
\caption{Anomalies in  $B$ meson decays. For the experimental values, statistical and systematic uncertainties are shown separately.
For $R_{K^{*}}$, we report in brackets the two LHCb measurements, in 
the $q^2$ bins $[0.045,1.1]\gev^2$ and $[1.1,6]\gev^2$. For \rdds, we use the 2017 HFLAV world average, which is based on \cite{Lees:2013uzd,Huschle:2015rga,Aaij:2015yra,Hirose:2016wfn,Aaij:2017deq}. For \rds, note that recent conservative SM estimates put the discrepancy at a slightly lower significance, see Ref.~\cite{Bigi:2017jbd}.
}
\label{tab:allB}
\end{table*}%

Such measurements are often presented as the ratios of
two related SM processes; while the SM prediction for individual branching ratios can have potentially 
large QCD uncertainties, they 
should cancel when taking the ratio between branching ratios into final states with different leptons \cite{Hiller:2003js}. We 
show in Table~\ref{tab:allB} the definitions of the various 
anomalous $R_X$ ratios, with $X=K^{(*)}, D^{(*)}, J/\psi$, together with the most up-to-date experimental 
measurements as well as the SM predictions. The discrepancies in both the charged current and neutral 
current interactions are each at the level of $4\sigma$. Other discrepancies exist in variables related 
to the angular distributions of $B$ decays, for example the $P_5'$ observable in $B^{0*}\to K^{0*}\mu^+\mu^-$ \cite{DescotesGenon:2012zf} displays an equally significant $2.6\sigma$ excess \cite{Aaij:2015oid,Wehle:2016yoi}.

Each set of discrepancies can be fit well by the addition of a single new physics operator. For example, for 
the anomalies concerning the $b\to s\mu^+\mu-$ transition one can add:
\beq
O={1\over \Lambda^2}(\bar \mu_L\gamma^\alpha\mu_L)(\bar s_L\gamma_\alpha b_L)
\eeq
where $\Lambda^2\sim 10^3 \tev^2$. Similarly for the $b\to c\tau \nu$ anomalies, the 
operator $(\bar c_L \gamma^\alpha b_L)(\bar \tau_L \gamma_\alpha \nu_{\tau L}) $ provides a good fit, where 
the scale $\Lambda^2\sim 1\tev^2$. Other choices of operators are possible, see for example \cite{Descotes-Genon:2015uva,Becirevic:2015asa,Hurth:2016fbr,Altmannshofer:2017yso,Bardhan:2016uhr,Capdevila:2017bsm,Hiller:2017bzc,Buttazzo:2017ixm,Alok:2017qsi}, with limits from high-$p_T$ LHC processes such as $p p \to \mu^+\mu^-$ or $p p \to \tau \nu$ still consistent with the scale necessary to explain the $B$ anomalies \cite{Faroughy:2016osc}.

Such operators are naturally generated by the exchange of a $Z'$ or a $W'$, or 
alternatively (after a Fierz rearrangement) by a vector leptoquark coupling to left-handed 
SM fields while violating flavor universality (see e.g. \cite{Freytsis:2015qca,Dorsner:2016wpm,Becirevic:2016yqi,Crivellin:2017zlb,Calibbi:2017qbu,Biswas:2018jun}).
The leptoquark alternative, however, often faces problems. Scalar and vector
leptoquarks can often have renormalizable or dimension 5 couplings that 
break $B$ and $L$, leading to 
unacceptably fast proton decay \cite{Dorsner:2012nq,Assad:2017iib}, although see \cite{Becirevic:2017jtw} for an exception. Furthermore, 
new scalars bring their own naturalness issues.

It is therefore natural to consider composite leptoquark models \cite{Gripaios:2009dq,Barbieri:2016las}, where 
the scalar
leptoquarks
are composite bound states of fundamental fermion fields that interact with quarks and leptons.
In these models, the naturalness issues are solved as in technicolor models,
while the operators inducing proton decay are also suppressed by the compositeness scale (however, TeV-scale composites require additional symmetries to satisfy proton decay bounds).
These and other motivations have led to many composite leptoquark models being proposed
\cite{Gripaios:2014tna,Cacciapaglia:2015eqa,Blanke:2018sro}.

Now, in any such model, there is typically a large spectrum of
composite objects. Over and above the leptoquarks themselves,
which are in the fundamental of color SU(3), a typical model contains composites
in the adjoint, sextet or singlet representations of color. The precise 
 spectrum  depends on the hidden gauge group and the
representations of the constituents.

In this note, we point out that this plethora of states is very relevant for the 
collider phenomenology of leptoquarks.
They provide potential new decay channels for the leptoquark, 
which can affect the phenomenology of leptoquark searches.
The usual searches assume that leptoquarks are
pair produced and simply decay to a quark and lepton,
thereby leading to a two lepton and two jet final state
(naturally, the precise signature depends on the quark and lepton flavor
that the leptoquark couples to, see e.g. \cite{Diaz:2017lit}).

On one hand, the new decay modes of the leptoquark 
can reduce the branching ratio into the usual channel and 
reduce the effectiveness of the search. On the other hand, the new decay modes open up 
new signals of leptoquarks. Furthermore, the additional states also
leads to
new production mechanisms of leptoquarks, through cascade decays of 
other composite states. 

These processes can significantly alter bounds on
leptoquarks, and also provide new signal regions to search for these models.
In all these cases, the
current collider 
bounds on the leptoquarks need to be reanalyzed. In principle, the bounds
 could be significantly weakened. If so, this could 
 open up the allowed parameter space of leptoquarks. In this
 work, we will explore this new set of signatures.

We note that new decay modes have been frequently considered in the context of leptoquarks (e.g.
\cite{Queiroz:2014pra}).
However, these often involve new particles put in by hand, and we have not noted any 
work (apart from the recent analysis in \cite{Chala:2018igk}) which uses the other states which are naturally produced by 
compositeness models.

\section{Models}

Composite models of leptoquarks
typically involve a new gauge group $G$, and 
fields in various  representations of the new group $G$.
One or more of these fields in addition carries 
color; these can then form bound states which 
could have the quantum number of leptoquarks.

{\it Model 1}. We shall take as our starting point the
 composite model for leptoquarks  presented in \cite{Cline:2017aed}. In the specific 
 model of \cite{Cline:2017aed}, 
the gauge group $G$ is taken  to be $SU(N)$.
In addition the model contains fields in the fundamental of this gauge group, shown in the following table:

\begin{center}
  \begin{tabular}{ | c || c | c |c |}
    \hline
     & $SU(N)$ & $SU(3)_c$ & $SU(2)_L$\\ \hline \hline
    $S$ & $N$ & 1 & 2\\ \hline
    $\psi$ & $N$ & 3 & 1 \\\hline
		    $\bar{S}$ & $\bar{N}$ & 1 & 2\\ \hline
    $\bar{\psi}$ & $\bar{N}$ & $\bar{3}$ & 1 \\\hline
  \end{tabular}
\end{center}

In addition \cite{Cline:2017aed} introduced a scalar $\phi$ with quantum numbers (1,2) under the SM gauge group.
This is required to allow the mixing between the composite leptoquark and the quarks and leptons of
the Standard Model using renormalizable operators. We 
will simplify the model by taking the $\phi$ field to be heavy,
and integrating it out. We will then work with 
the effective interaction \bea
{\cal L}_{eff}=\lambda\bar{Q}L\psi\bar{S}+h.c.\label{interaction}
\eea

When the gauge group $SU(N)$ confines, we get a spectrum of bound states.
In particular, the 
bound state of $\bar\psi$ and $S$ has the quantum numbers of  a composite leptoquark. 
In addition to this leptoquark,
 there
are two other mesonic bound states: the $\psi\bar{\psi}$ and $S\bar{S}$
bound states. 
The $S\bar{S}$ is neutral under color, while  the  bound state $\psi\bar{\psi}$ 
yields a neutral state and a state in the adjoint of color.\footnote{Ref.~\cite{Cline:2017aed} also considered a SM singlet $S^N$ which can serve as a dark matter candidate. It will not play a role in the phenomenology described here.}

{\it Model 2.} We can also consider more complicated
models for composite leptoquarks.
 For example, consider  a vector-like model with a 
scalar field $A$ in the antisymmetric tensor representation,
 in addition to the  $S, \psi$ fields
of the above model. 
\begin{center}
  \begin{tabular}{ | c || c | c |c |}
    \hline
     & $SU(N)$ & $SU(3)_c$ & $SU(2)_L$\\ \hline \hline
		    $A$ & { asym} & 1 & 2\\ \hline
						    $\bar{A}$ &  $\overline{\rm asym}$ & 1 & 2\\ \hline
    $S$ & $N$ & 1 & 2\\ \hline
    $\psi$ & $N$ & 3 & 1 \\\hline
			    $\bar{S}$ & $\bar{N}$ & 1 & 2\\ \hline
   $\bar{\psi}$ & $\bar{N}$ & $\bar{3}$ & 1 \\\hline
  \end{tabular}
\end{center}

In this model,
 there are  additional bound states involving the antisymmetric.
There is a bound state of
 $A$ with two $\psi$;
this is in the sextet of color, and has lepton number 2. Furthermore, bound states of
$\bar{A}$ with $\psi S$ and $\psi \phi$ produce additional leptoquarks.

Furthermore,
we can add the interaction
\bea
{\cal L}=\lambda_4 A_{ij}\bar{S}_i\bar{S}_i
\eea
Note that this preserves lepton number if we 
 assign $A$ a lepton number of 2.

\section{Composite Leptoquark Decay Modes}

As we have described, the composite leptoquark models
have several bound states. The phenomenology of these models 
then depends on the spectrum and interactions of the bound states.

The spectrum cannot always be exactly evaluated, due to the strongly coupled
nature of the bound states. To explore the possibilities, we will always 
assume the confinement scale $\Lambda$ to be lower than the constituents masses
and therefore focus on the phenomenology of a not very strongly coupled theory, 
where we expect the masses of the composites to track the masses of the constituent fields.
We can therefore order the bound state masses by
choosing the ordering of the masses of the constituent fields.
We do this to paint a picture of the different resulting simplified models whose
LHC phenomenology is studied, and expect the spectrum in the full theory to
get $O(1)$ corrections due to the confining dynamics.

For example, we can take the 
mass of $\psi$ (which is the same as the mass of $\bar{\psi}$) to be smaller  than  that of $S$. This
is expected to lead to an ordering of bound state masses such that  $m(S\bar{S})>
m(\psi\bar{S})=m(S\bar{\psi})> m(\psi\bar{\psi})$. 
There is a further splitting between the adjoint and the scalar of $\psi\bar\psi$ due to QCD effects which
we shall ignore in this paper; it may however be important in the
highly degenerate case. 
In this ordering, pair-produced $\psi S$ leptoquarks can decay to a quark, a lepton and $\psi\bar\psi$, which then decays to jets. 

Alternatively we can take the 
$\psi$ mass to be larger
  than  that of $S$. This
is expected to lead to an ordering of masses such that  $m(\psi\bar{\psi}) >
m(\psi\bar{S})=m(S\bar{\psi})> m(S\bar{S})$.
This is a more interesting situation because the decay of the leptoquark
to the $S\bar S$ state (along with a quark and a  lepton)
may partially hide the leptoquark. We will focus on this
ordering for the rest of this paper, leaving the case discussed in the previous
paragraph to future work.

The next piece required for the phenomenology is to
identify the dominant decay mode of the $S\bar{S}$ bound state.
In this model, it is likely that the dominant mode
is to two quarks (through an off-shell $\psi$ and $\bar{\psi}$) 
and we shall assume that this decay occurs 100\% of
the time.
We assume that the alternative decay to leptons is suppressed.
It may be interesting to consider subdominant decay modes
such as to two photons; we shall again not consider this further here.

The final piece for leptoquark decay is 
to determine the branching ratio for the two
possible decay modes $\psi\bar{S}\to l+q$
and $\psi\bar{S}\to S\bar{S}+l+q$. 
Both these decays occur through a single insertion
of the operator (\ref{interaction}) that allows  $\psi$  to decay to
 $S$ plus  a quark and lepton. However, the first process requires the
 $\psi$ and $S$ to be at the same point, and the rate is therefore
 suppressed by a factor $|\psi(0)|^2$, where $\psi(0)$ is the wavefunction
 at the origin. 
 At least for weak coupling, this leads to a suppression by a factor $g^3$
 where $g$ is the coupling of the $SU(N)$ gauge group at the scale of the
  wavefunction.
 If the bare field masses are much larger than the confinement scale 
 (as assumed above), this suppression can be significant. For example, Ref.~\cite{Cline:2017aed} finds
typical values for $|\psi(0)|^2/M^3~10^{-3}$, where $M\sim1$ TeV are the constituent masses. In the next sections,
 we will discuss the LHC coverage of both cases of dominant exotic decays of 
 the leptoquark, as well as mixed exotic and standard decays.
 
 Additional states can arise from confining $\psi, S$ with other fields charged under the confining group. For example, the scalar doublet $\phi$ of \cite{Cline:2017aed} might not be too heavy and it might not be appropriate to integrate it out to get the effective interaction \eqref{interaction}: then, fermionic bound states with the quantum numbers of quark and lepton  $(\bar \psi \phi$ and $\bar S \phi)$ would be present in the low-energy theory. We neglect those and refer to the recent work \cite{Chala:2018igk} for an LHC analysis of the lepton-like composites.
 
 With these assumptions, we find that the pair produced 
 leptoquarks dominantly decay to two leptons and 6 jets, with the leptons potentially soft. This is significantly different from the usual leptoquark
 signature.

\section{Leptoquark  Production in Composite Models}

The new composite states  can also provide new
production modes of the leptoquark.
 As we have already seen, composite models
 generically contain bound states 
in large representations of 
color. In that case, they can be produced more copiously
than the leptoquarks even if they are more massive, with the color factor outweighing the
phase space suppression.
Their decay can potentially provide a new production mode of leptoquarks.

 We note that each of the $\psi\bar\psi$ 
states can decay to a leptoquark and a quark and lepton through an insertion of
the operator (\ref{interaction}).
The further decay of the 
leptoquark and the $S\bar{S}$ states
 then produces a signal of eight jets and and four leptons; which is
once again quite different from the usual leptoquark signature.

An even more striking process 
may be generated in the second model above.
In this model, we may take
the  mass of the antisymmetric tensor to be larger than
twice the $S$ mass i.e. $m_A>2m_S$. In this case,
the $A \psi\psi$ bound state  can 
decay to two $\psi\bar{S}$ leptoquarks.
This can lead to a striking signal where
the sextets can be pair produced and then 
decay to four (boosted) leptoquarks.
The final signal then can 
have four leptons, accompanied by as many as twelve jets.
This 
explosive signal can be a striking signal of this model.

\section{A simplified model}
\label{sec:simpmod}
To study these multitude of possibilities, we will use the language
of simplified models.
Motivated by the above models,
we  shall consider a simplified model with three new BSM particles;
a leptoquark  which we denote $\Phi_{LQ}$, a neutral state
$N$, and a colored state $\Phi_{C}$. There are two models:
in one the colored state has lepton number 0, and in the other it has lepton number 2.
We shall assume a mass ordering where
$m(\Phi_{C})>m(\Phi_{LQ})> m(N)$ and further assume that the neutral state $N$ decays promptly to two jets.

We shall introduce one interaction leading to decay of the
leptoquark to the neutral state plus a quark and lepton
\bea\label{eq:LagLQ}
{\cal L}\supseteq\Phi_{LQ}N \bar{Q}L
\eea
In the following we assume that this interaction dominates the leptoquark decays while the 
two-body branching ratio is assumed to be negligible.

In the first model, where the colored state has lepton number zero, the colored state will 
be taken to decay promptly to the leptoquark
through an interaction
\bea\label{eq:LagOctet}
{\cal L}_8=\Phi_{LQ}\Phi_C \bar{Q}L
\eea

On the other hand, if the colored state has lepton number 2, we will consider the 
interaction
\bea\label{eq:LagSextet}
{\cal L}_6=\Phi_{LQ}\Phi_C \Phi_{LQ}
\eea
which allows the colored state to decay to two leptoquarks.

As discussed above, the composite state masses originate both from the confining 
dynamic as well as the constituent bare masses. We shall not impose any theoretical prejudice,
and we will analyze the simplified models allowing the various particles to have any possible mass. 

We dsicuss flavor constraints for this model, as one might expect them to exclude some of the parameter space, especially for sub-TeV composites. For our reference model of Eq.~\eqref{interaction} this was considered in Ref.~\cite{Cline:2017aed}, where it was found that TeV scale composites can solve the $B$ flavor anomalies and at the same time be compatible with meson mixing (induced by the octet) and $f\to f'\gamma$ transitions (induced by the heavy fermions $Q\phi$ and $L\phi$). In particular, the mixing limits from $K^0, D^0$ and $B_s^0$ were all found to be almost saturated: while the author was mostly considering  the case of near-degenerate composites, we will see that with the additional interactions the direct searches on the scalar octet considerably are exclude higher masses, thus reducing this tension. Finally, the interactions introduced above will not generate large flavor violation as any induced process would have higher loop order. As the sextet carries both color and lepton number, it does not couples directly to quarks and leptons and therefore its contribution to any flavor violating process will only come at higher loop level.

Note that we have not specified the quark and lepton flavors in the Lagrangians above: in the rest of this 
work we will always assume that one specific flavor combination dominates the couplings, and we will either 
take $q\ell=c\tau$ or $q\ell=b\mu$ as benchmarks that can respectively be used as explanations of 
the $R_D$ or $R_K$ anomalies.\footnote{%
The anomalies are better fit by a vector 
leptoquark instead of a scalar. We will here restrict ourself to the collider implications of 
the scalars, because the vector coupling to gluons, and therefore the production cross section, depends on the UV dynamics. When $SU(N)$ confines, the 
low-energy spectrum will include both (pseudo)scalar and (axial)vector mesons, so it can be expected 
that, in the case $B$ anomalies are mediated by the vector meson, the scalar is at a comparable scale and is also accessible at the LHC.}
Other choices will yield limits between the two cases that we present: when $q\ell=b\tau$, the higher $b$-tagging efficiency will increase limits with respect to $c\tau$, while for $q\ell=s\mu$ the lack of $b$'s will result in slightly lower limits compared to $b\mu$. Finally if the lepton is a neutrino, standard SUSY searches exclude colored scalars decaying to a jet and massless missing energy up to 1\tev.

\section{Searches for composite leptoquarks}

\begin{table}[tb]
\begin{center}
\begin{tabular}{c|c|c}
Search & Reference & $\mathcal{L}_{int}$ (\ifb) \\\hline
CMS 3rd gen. LQ & CMS-EXO-16-023 \cite{Khachatryan:2016jqo} & 12.9  \\
%CMS 2nd gen. LQ & CMS-EXO-16-007 \cite{Sirunyan:2017yrk}  & 2.7 \\
CMS 2nd gen. LQ & CMS-EXO-17-003 \cite{CMS:2018sgp}  & 35.9 \\
ATLAS RPV 1L  &  ATLAS-SUSY-2016-11 \cite{Aaboud:2017faq} & 36.1   \\
CMS RPV 1L &  CMS-SUS-16-040 \cite{Sirunyan:2017dhe}  & 35.9 \\
\end{tabular}
\end{center}
\caption{Searches recasted in this work. All are based on the $\sqrt s=13\tev$ dataset.}
\label{tab:recasted_searches}
\end{table}%

While ATLAS and CMS have not specifically targeted the models discussed here so far, our final states have at least two leptons and multiple jets, with little missing energy. There are
hence many possible searches with final states leptons which are sensitive to these
models.
We here list the LHC searches that we used in this work to constrain the leptoquark models
(these are also summarized in 
Table \ref{tab:recasted_searches}), along with a few relevant details.

We shall now analyze these models as a function of the masses of the various particles, and set limits by 
recasting the relevant existing searches from ATLAS and CMS.

\paragraph{\it Second generation leptoquark searches:} 
The CMS search for second generation leptoquark \cite{CMS:2018sgp} targets 
leptoquark pair-production, followed by the decay $\Phi_{LQ}>j\mu$. Inclusive signal regions are 
defined at thresholds of $S_T$ (the scalar sum of the transverse momentum of the two leading 
muons and jets), $M_{\mu\mu}$ (the invariant mass of the dimuon pair) and $M_{min}(\mu,j)$, the 
smaller of the muon-jet invariant masses that minimizes the mass difference between the two 
reconstructed leptoquarks. This analysis has recently been released with the full 35.9~\ifb\ of the 2016 dataset.\footnote{The first preprint version of this work showed limits from the previous CMS leptoquark search, based on only 2.9~\ifb, which resulted in far weaker limits.} A similar search 
by ATLAS \cite{Aaboud:2016qeg} is based on only 3.2~\ifb and results in weaker 
limits: we therefore do not reinterpret it.
\paragraph 
{\it Third generation leptoquark searches:}
Here the most stringent analysis is 
the CMS search for third generation leptoquark \cite{Khachatryan:2016jqo} targeting leptoquark 
pair-production, followed by the decay $\Phi_{LQ}>b\tau$, where one $\tau$ lepton decays 
hadronically and the other one leptonically.  Events with at 
least one $b$-jet are selected and exclusive bins are defined with respect to the variable $S_T$, defined 
as the scalar sum of the transverse momentum of the hadronic tau, the identified lepton, and the two 
leading jets, together with the transverse missing momentum. The analysis is based on 12.9~\ifb of data at 13~TeV.
\paragraph {\it R-parity violating supersymmetry searches:} 
If R-parity is violated, all supersymmetric particles 
are unstable, and the signals have no missing energy but larger jet multiplicities. 
The final states then resemble the ones considered here.
The ATLAS 
search for RPV SUSY \cite{Aaboud:2017faq} requires one hard ($p_T>30\gev$) lepton and between 8 and 12 
jets above $p_T$ thresholds of 40, 60 and 80\gev, with model-independent inclusive signal regions 
defined for different $n_{jet}$ and $n_{b-jet}$ multiplicities. This search is based on the full 
public 13 TeV dataset from ATLAS, 36.1~\ifb.
CMS also has a similar RPV search targeting final states with one lepton and 
multiple jets \cite{Sirunyan:2017dhe}. 
For the models considered in this work, limits from this search were found to be 
always subdominant with respect to the corresponding ATLAS analysis.
\paragraph {\it Dijet resonance searches:} 
If the leptoquark and the neutral state are close in mass, the emitted leptons can be very soft.
The model is still constrained by the dijets coming from the decay of the neutral state.
The ATLAS paired 
dijet resonance search \cite{Aaboud:2017nmi} looks for a bump in the dijet invariant 
mass distribution in events with four jets, in order to find pair-produced particles decaying 
to dijets. This search sets a limit on 450\gev\ on RPV stops, which we will apply to our model in 
degenerate regions of the parameter space.

Each analysis is validated against the simplified models considered in the original paper, for which we 
reproduce the exclusion limits. We generate events with \texttt{MadGraph\_AMC@NLO 2.5.3} \cite{Alwall:2014hca}, 
shower and hadronize them with \texttt{Pythia 8.219} \cite{Sjostrand:2014zea}, and 
use \texttt{Delphes 3.4.0} \cite{deFavereau:2013fsa} for detector simulation, for which we use 
the ATLAS and CMS detector cards developed in \cite{Asadi:2017qon}. Jets are clustered in \texttt{FastJet 3.2.1} \cite{Cacciari:2011ma} with the anti-$k_T$ algorithm \cite{Cacciari:2008gp} and a jet radius of $0.4$. Finally, we simulate the 
particle event selection and cutflow of the analyses with \texttt{ROOT}. 
Source code and validation material for each analysis are provided as additional material in the 
arXiv submission.

To set limits on BSM physics we use the profile likelihood ratio test statistics, where we minimize the Poisson likelihood with respect to Gaussian nuisance parameters representing background 
uncertainties in each bin. The ratio is taken between the likelihoods given the two
 hypotheses of background vs signal plus 
background (we use the Python package \texttt{iMinuit} to numerically minimize the 
likelihood function with respect to the nuisance parameters). For more details, 
see \cite{Cowan:2010js} or Appendix A of \cite{Asadi:2017qon}. In searches where the bins are exclusive, we use the combined event counts from all bins to construct the likelihood (neglecting possible correlations between bins, as they are not made public for the searches we are using), while for inclusive bins, limits are set using the signal region with the best expected exclusion reach.

\begin{figure}[t]
\begin{center}
\includegraphics[width=0.75\columnwidth]{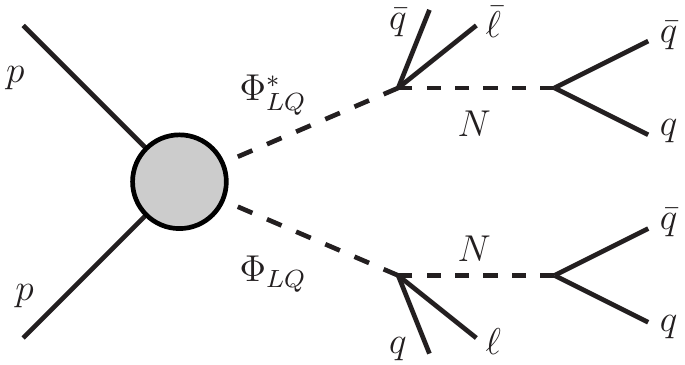}
\caption{Leptoquark production and decay chains considered.
}
\label{fig:decays}
\end{center}
\end{figure}

\section{Limits}

In the following, we discuss limits for the simplified model framework 
outlined in Section~\ref{sec:simpmod}. As discussed above, we will
consider two cases for the quark and lepton flavors
in the decay, where the quark and lepton are either $b\mu$ or
$c\tau$.
We also assume all decays to be prompt.

We first consider QCD pair-production of the colored 
leptoquark $\Phi_{LQ}$, followed by the decay to the neutral scalar $N$ as in 
Fig.~\ref{fig:decays}. We then scan the $m_{LQ}-m_N$ plane.

We start by assuming 100\% branching ratios into the specified decay 
chain. 
\begin{figure}[t]
\begin{center}
\subfigure[\ $c\tau$-dominated final state.]{\label{fig:limits1}
	\includegraphics[width=\columnwidth]{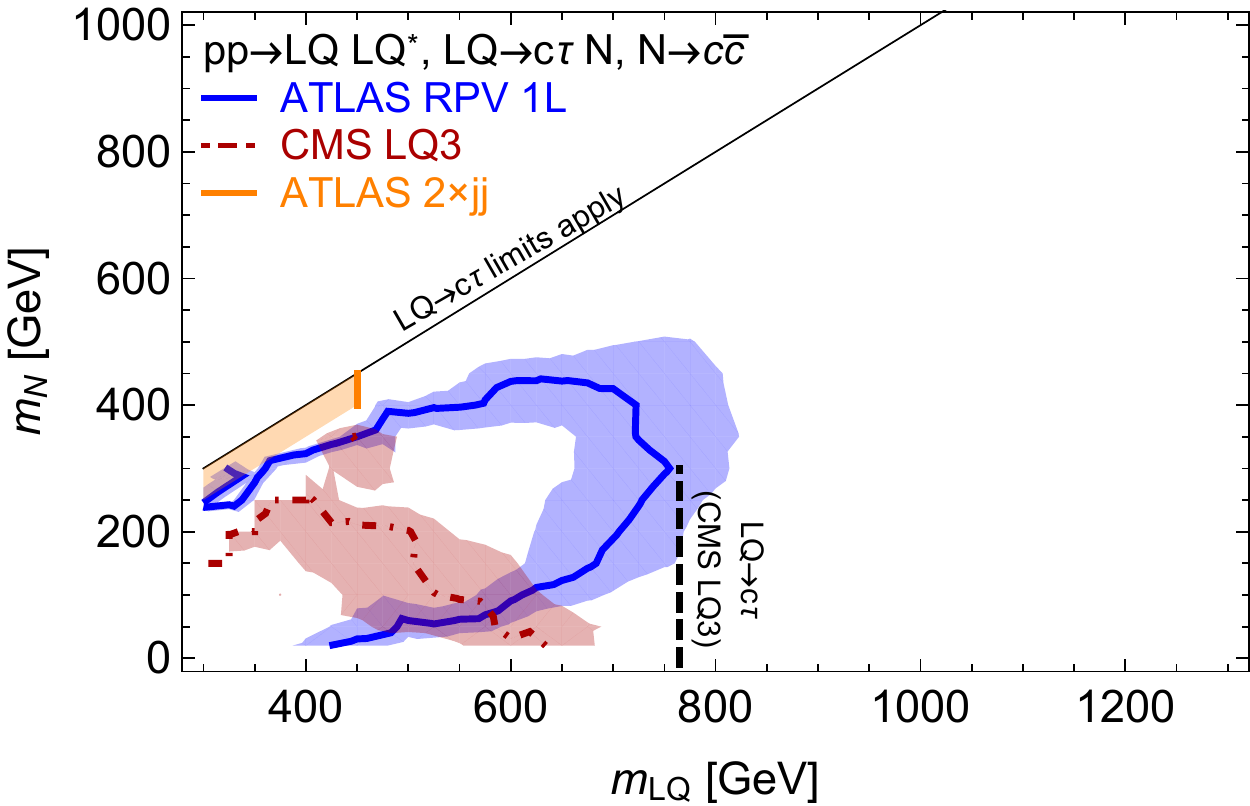}}
\subfigure[\ $b\mu$-dominated final state.]{\label{fig:limits2}
	\includegraphics[width=\columnwidth]{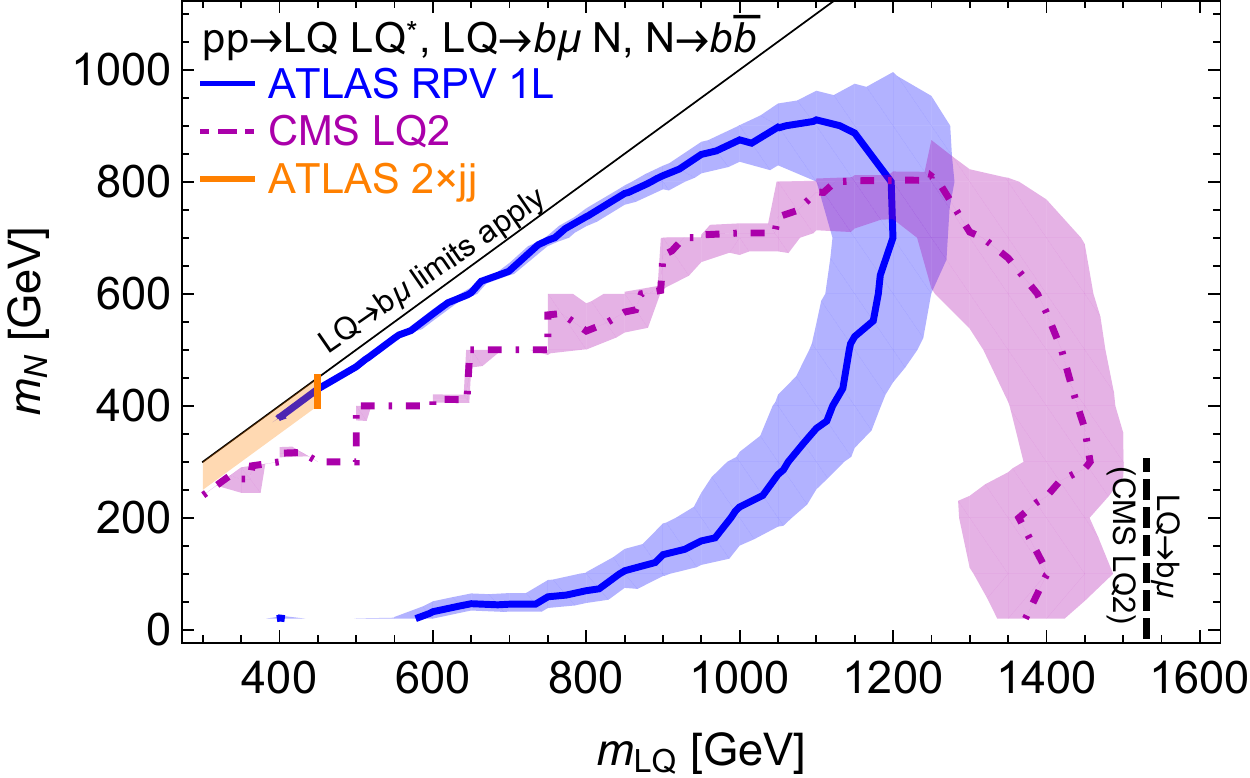}}
\caption{Recasted limits on leptoquark pair-production, followed by the decay $\Phi_{LQ}\to q\ell N, N\to \bar qq$. 
In each case, the vertical dashed line shows the nominal limit for the two-body decay $\Phi_{LQ}\to q\ell$, with branching ratio equal to one. Note the larger mass range displayed for $q\ell = b\mu$.
}
\label{fig:limits}
\end{center}
\end{figure}
The resulting exclusions are depicted in Fig.~\ref{fig:limits}. Here, limits are shown as lines (areas to the left  of each curve are excluded) with shaded bands depicting the range in which limits change when the number of signal events is augmented or reduced by a factor of $1.5$. The bands reflect the inherent uncertainty present in any external reinterpretation of LHC analyses, stemming from a crude detector simulation and treatment of object reconstruction efficiencies (e.g. for lepton identification, or $b$- and $c$-tagging). Furthermore, it easily allows us to identify the amount by which the limits weaken if the branching ratio into the final state studied is smaller than one. 

In Fig.~\ref{fig:limits1}, we show the limits for the decays with charm quarks and tau leptons, while the case of bottoms and muons is below in Fig.~\ref{fig:limits2}. 
In most of the parameter space, the most constraining analysis is the ATLAS RPV search (solid blue line). When $\ell=\tau$, this search relies on the leptonic decay mode of at least one of the two taus. At low masses ($m_{LQ}\lesssim 700\gev$), limits are mostly set by the signal region with at least eight jets with $p_T>60\gev$ and zero (three) $b$-jets for the $q=c$ ($q=b$) case. At larger masses, the signal region with at least eight jets with $p_T>80\gev$ becomes dominant. When the leptoquark and the neutral scalar are nearly degenerate, the lepton in the first step of the decay chain becomes soft and the event would not pass the one-lepton trigger. This is more evident when $\ell=\tau$, as the small tau transverse momentum is shared between the lepton and two neutrinos.

In the other limit, when the neutral scalar mass $m_N$ becomes small, the limits weaken as the $N$ is more boosted and its decay products overlap, resulting in fewer reconstructed jets in the final state. Notably, the second- and third-generation leptoquark searches set the strongest limits in this last case (dot-dashed lines). In this part of parameter space, the final state resembles the most to the standard two-body decay. As the searches only consider the leptons and the two hardest jets to try to reconstruct the leptoquarks (therefore missing the remaining jet), the efficiencies will be lower than in the two-body case. We note that in Fig.~\ref{fig:limits1}, the required $b$ jet arises from mistagging one of the $c$ jets (the mistag rate we use is of order $15-20\%$  in the  $ 50-500 \gev$  $p_T$ range), while in Fig.~\ref{fig:limits2}, the presence of six $b$ quarks in the event is irrelevant as the second-generation leptoquark search does not use $b$-tags.

Finally we also show the nominal limit from the ATLAS paired dijet search \cite{Aaboud:2017nmi} near the degeneracy line (orange shading) up to 450\gev.

In both cases, for reference we also show the nominal limit on the leptoquark in the standard (although family-violating) two-body decay $\Phi_{LQ}\to q\ell$  (dashed vertical line). Comparing to the three-body decay limits, one can judge if and where the additional dynamic opens up the allowed parameter space.
It can be seen that, apart from ``optimal'' regions of parameter space with moderate mass splittings, longer decay chains generally weaken the limits. This is particularly true for nearly degenerate masses and for final states with tau leptons (this could be expected, as efficiencies for hadronic tau tagging are only of order 60\%), for which leptoquarks could still be as light as $450-500\gev$, with mass splittings of order 100\gev.

Our approach so far has been to assume 100\% branching ratio for $\Phi_{LQ}\to q\ell N$. It should be noted that for a  model with given couplings for the interaction terms \eqref{interaction} and \eqref{eq:LagLQ} in the Lagrangian, the two-body decay $\Phi_{LQ}\to q\ell$ becomes favorable as $m_{LQ}$ gets closer to $m_N$ and the three-body decay is phase-space suppressed. Therefore, limits from standard leptoquark topologies would become relevant and can exclude the degenerate region.

\begin{figure}[t]
\begin{center}
\subfigure[\ $c\tau$-dominated final state.]{\label{fig:limitsBR1}
	\includegraphics[width=\columnwidth]{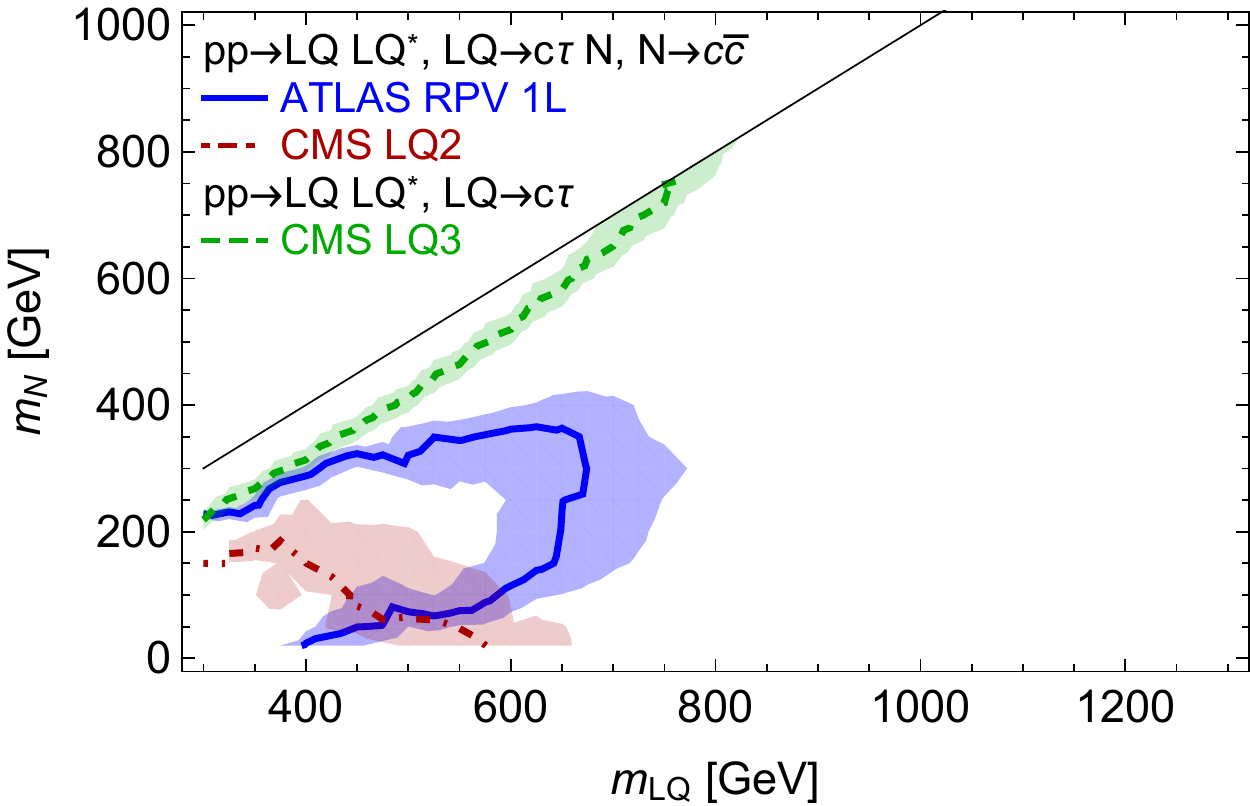}}
\subfigure[\ $b\mu$-dominated final state.]{\label{fig:limitsBR2}
	\includegraphics[width=\columnwidth]{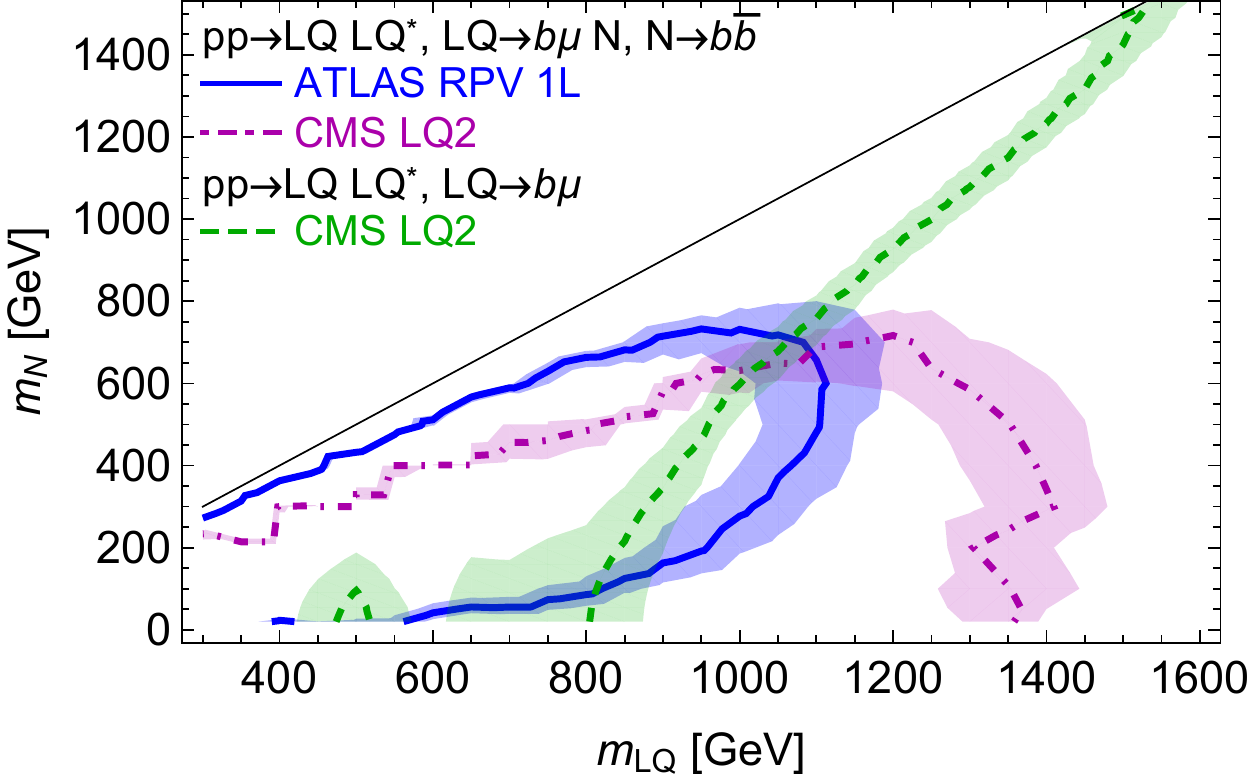}}
\caption{Same as Fig.~\ref{fig:limits}, but starting with a two-body branching ratio of 10\% at low $m_N$ and taking into account the phase-space suppression of the three-body decay near the degeneracy. The dashed line shows the limit from the two-body decay mode $\Phi_{LQ}\to q\ell$.}
\label{fig:limitsBR}
\end{center}
\end{figure}

We display this behavior in Fig.~\ref{fig:limitsBR}, where we simply set parameters such that $Br_0({q\ell})=10\%$ in the low $m_N$ limit. Then, as $m_N$ increases, the three-body decay phase space is suppressed and the limits from the standard leptoquark decay become more important. Note that this branching ratio was only taken as an example: higher or much lower values might be relevant in the composite model (as an example, in \cite{Cline:2017aed} the pseudoscalar coupling to SM fermions is helicity suppressed). Still, for $\ell=\mu$, the nearly degenerate region at $m_{LQ}<1~\tev$ becomes accessible only for parameters such that $Br_0({q\ell})<1\%$.

We conclude this section by noting that the weaker limits on final states with taus are not only due to the difficulty with tagging hadronic tau decays, but is mostly to the absence of an optimized search for this final state. In fact, no $\tau+$jets search exist at present (we will discuss a possible extension of the current lepton+jets searches to cover this possibility in the next section).

\subsection{Projections}
The experimental searches used in this work were based on the currently available 13\tev\ dataset, and in some cases on a small subset of the $2015-2016$ runs. With the 2017 run having already collected an additional 50\ \ifb, and the 2018 run about to start, we can ask how much of the parameter space can be covered in the near future. For definiteness, we take as a target the 300\ \ifb\ that should be collected before the start of the HL-LHC run.

To get a qualitative estimate of the future reach, we simply scale the excluded cross section by the relative $\sqrt{\mathcal{L}_{int}}$ increase for each search. This is a conservative estimate that can be reached by simply collecting more data in the same bins as in the existing searches (when statistics-dominated), while experimental collaboration will likely be able to define finer bins as more data is collected and therefore achieve better discriminatory power. 

\begin{figure}[tb]
\begin{center}
\includegraphics[width=\columnwidth]{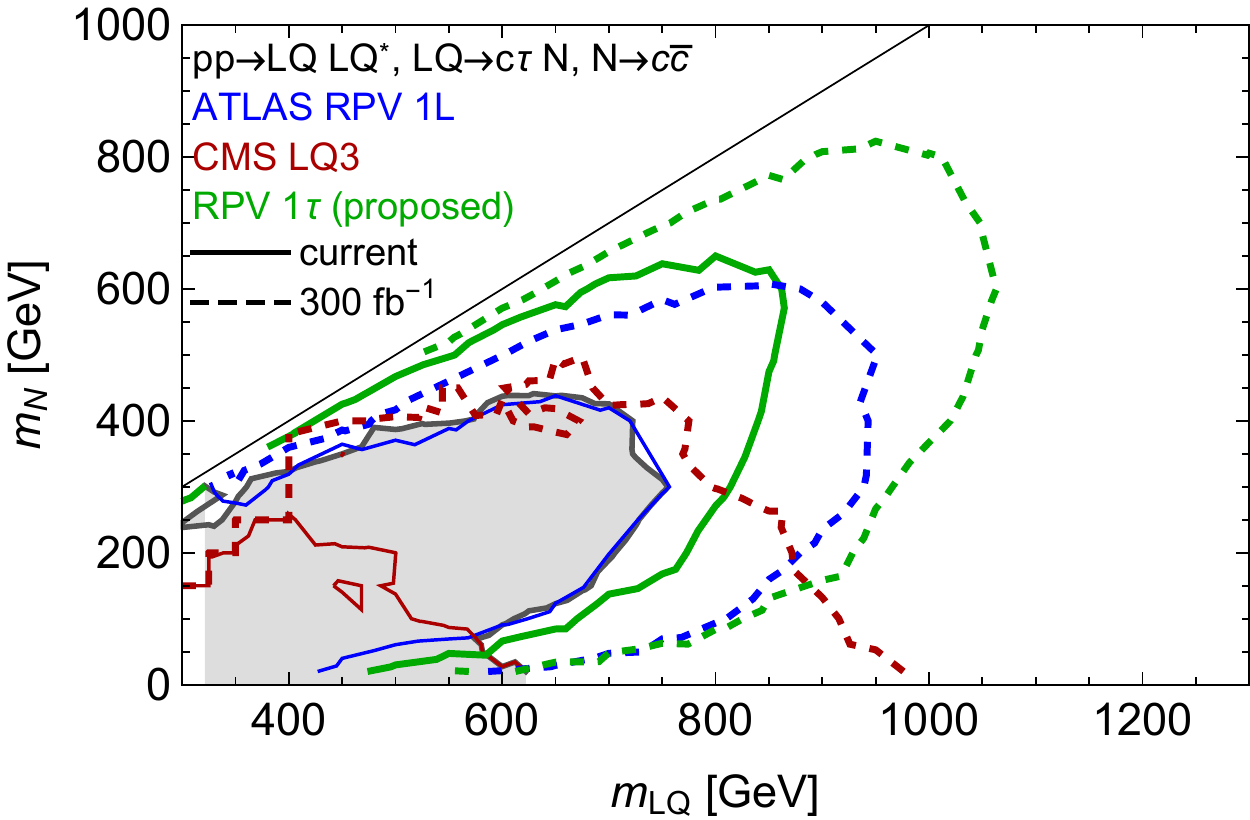}
\caption{Reach of a $\tau$+jets search and projected limits on leptoquarks with a $c\tau$-dominated final state for an integrated luminosity of 300 \ifb (Dashed lines).
}
\label{fig:limits_proj}
\end{center}
\end{figure}

We show the projected reach in Fig.~\ref{fig:limits_proj} for the $c\tau$-dominated final state: the shaded gray area shows the current nominal limits based on the current dataset, while the dashed lines show the prospected improvement for each search. In addition to the current existing searches, we also show (in green) the current and projected reach of a $\tau$+jets search: here, we simply take the signal region definitions of the ATLAS RPV 1L search \cite{Aaboud:2017faq}, but require a reconstructed hadronic tau instead of a muon or electron (we consider a tau tagging efficiency of $60\%$). We then assume that, if performed on the same dataset, the search can perform similarly to the ATLAS analysis, which yields limits of the order of $0.5-1$~fb for a BSM signal cross section in the relevant signal regions (varying the sensitivity of the search in this range would only raise or lower the maximum limits by about 50 GeV with respect to Fig.~\ref{fig:limits_proj}). It can be seen that a $\tau$+jets search based on $36~\ifb$ could already outperform all existing searches, both in the bulk and in almost-degenerate regions of parameter space.

For the other final state ($q\ell= b\mu$), the current limits already exclude a leptoquark below 1\tev (outside of the degenerate region). We calculate that with $300~\ifb$ of data, the reach will be $1350-1450\gev$ for the leptoquark and up to $1100\gev$ for the neutral state.

Especially with the addition of a new search focused on $\tau$'s, it can be seen that collecting more data can more than double the currently covered range of leptoquark masses.
This is in particular relevant if leptoquarks are an explanation of the $R_{D^{(*)}}$ anomaly, where the BSM Wilson coefficient is of order $(1\tev)^{-2}$. Barring large couplings to SM fields, this would imply that the LHC can cover much of the relevant parameter space.\footnote{
For couplings larger than one, additional limits arise from single leptoquark production $qg\to \Phi_{LQ}\ell$ \cite{Dorsner:2014axa,Dorsner:2018ynv,Hiller:2018wbv}, which was recently studied at CMS \cite{CMS:2018hjx}.}

\section{Limits on additional colored scalars}

We now discuss limits on the extended leptoquark sector, in which additional colored scalars $\Phi_C$ can be produced at the LHC. In particular, the additional scalars come in large $SU(3)_c$ representations, such as sextet or octet, which therefore result in higher cross sections. Note that in the simplified model language, there are now three independent mass scales (the masses of $\Phi_C, \Phi_{LQ}$ and $N$), while in a UV-complete model the masses could be related to each other. To simplify the parameter space, we consider the $m_{C}-m_N$ plane and assume that the leptoquark mass is in between.

In particular, we take $m_{LQ}=(m_C+m_N)/3$. We have verified by taking different {\it ansatze} and slices at constant values of the $N$ mass that the limits do not vary much with the intermediate leptoquark mass. Due to the large mass scales and high jet and lepton multiplicities, the signal populates enough bins independently of the intermediate state.

With the simplified Lagrangians in Eqs.~\eqref{eq:LagOctet} and \eqref{eq:LagSextet}, $\Phi_C$ pair-production results in a striking final state: eight quarks and four leptons in the first case, twelve jets and four leptons in the second. We again find that the ATLAS RPV search \cite{Aaboud:2017faq} provides the best coverage, with selection efficiencies of order $30-50\%$, largely independent of the masses (reduced by the leptonic $\tau$ branching ratio for decays with taus). The limits therefore scale mostly with the production cross section.

\begin{figure}[tb]
\begin{center}
\includegraphics[width=\columnwidth]{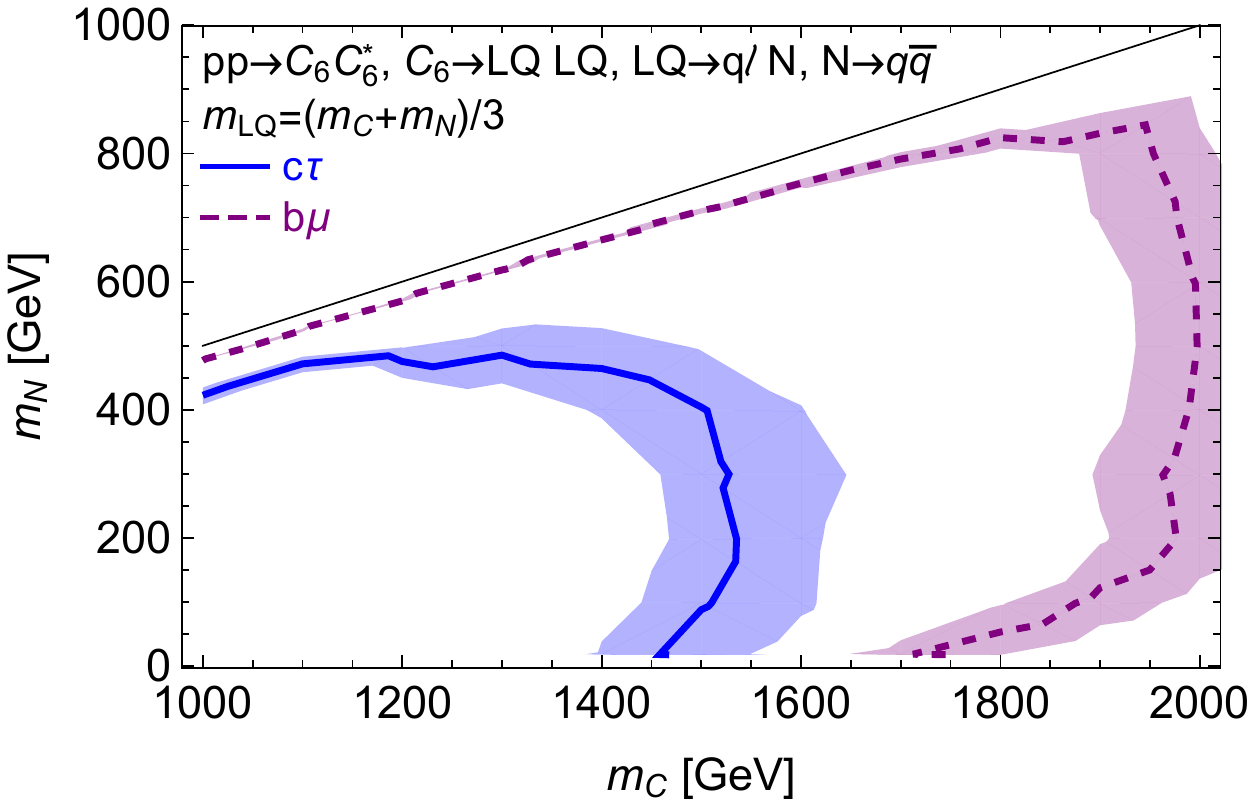}
\caption{Limits on colored (sextet) scalars decaying to two leptoquarks, for different flavor compositions of the final states. Limits on the octet model are comparable.
}
\label{fig:limits_color}
\end{center}
\end{figure}

We show the resulting limits in Fig.~\ref{fig:limits_color}. Again, we see that the absence of a dedicated search does not make the LHC blind to this signal. In fact we find limits on exotic scalars in the $1.4-1.5\tev$ range for $q\ell=c\tau$, and $1.8-2\tev$ for $q\ell=b\mu$. 
Note that the octet cross section is higher than for the sextet by a factor of $2$, while the efficiencies are slightly lower due to the lower jet multiplicity: we find that the resulting limits are approximately the same as in Fig.~\ref{fig:limits_color} (after rescaling $m_N\to 2 m_N$).
As before, near the degeneracy our assumption of neglecting other decay modes could break down: for example, the octet $\Phi_C=(\bar \psi \psi)$ can annihilate directly to gluons, while the sextet can decays to two quarks and two leptons.

Limits on the additional colored states can be compared to the direct limits from leptoquark pair-production discussed previously. In particular, the assumed decay mode relied on the mass ordering $m_{LQ}<m_C$ ($\tfrac12m_C$) in the octet (sextet) case. For example, the direct $m_{LQ}>600\gev$ limit for $m_N=50\gev$ and $c\tau$ automatically  requires $m_C>600\ (1200) \gev$ for this decay mode to be open. On the other hand, the direct limit of $m_C>1450\gev$ found here is stronger. Similar arguments apply to the degenerate as well as the $b\mu$ cases.

In this sense, more parameter space can be covered by looking at direct production of additional composite states.

\section{Conclusions}

In this work, we have studied some of the collider signatures of a composite leptoquark sector, as motivated by the numerous $B$ physics anomalies. We have pointed out that the standard leptoquark search channel $\Phi_{LQ}\to q\ell$ can overestimate the current reach of the LHC for a direct corroboration of the new physics possibly responsible for these anomalies. 

In particular, we have shown that composite leptoquarks related to the charged current anomalies (\rdds\ and \rjp) could still be as light as $500-600$\gev, with no particular tuning or mass degeneracy. On the other hand, limits on leptoquarks related to the neutral-current anomalies (\rkks, angular distributions) are more robust, with sub-TeV leptoquarks only allowed for relatively narrow mass splittings ($50-100\gev$). We have also shown that higher color representations provide spectacular collider signatures, and can translate into stronger limits on these exotic states.

If the hints of lepton-flavor non-universality in $B$ decays are to be confirmed, one can hope that colliders can directly produce the mediators. It is then important to thoroughly cover the possible detector signatures of such states.

\section*{Acknowledgements}

We thank Pouya Asadi, Matt Buckley, Anthony DiFranzo, and David Shih for their contributions in developing the code-base used in this paper. A. R.  was supported in part by NSF Grant No.~PHY-1620638.
 A. M. was supported in part by NSF Grant No.~PHY-1620638, and in part by Simons Investigator Award \#376204.

%\bibliography{biblio}

\end{document}